\documentclass{PoS}
\usepackage{amsmath}
\usepackage{mathtools}
\usepackage{amssymb}
\usepackage{graphicx}

\title{
Shannon entropy and hadronic decays
}

\ShortTitle{Shannon entropy and hadronic decays}

\author{\speaker{Felipe J. Llanes-Estrada}\thanks{Work supported by grants from MINECO FPA2011-27853-C02-01 and FPA2016-75654-C2-1-P, and carried out in the inspiring atmosphere of the theoretical physics department and UPARCOS.},\ \ Pedro Carrasco Mill\'an, Ana Porras Riojano and Esteban M. S\'anchez Garc\'{\i}a\\
Departamento de Fisica Teorica I, Plaza de las Ciencias 1, Fac. CC. Fisicas; Universidad Complutense de Madrid, 28040, Madrid, Spain.\\
        E-mail: \email{fllanes@fis.ucm.es}}
\author{
M. \'Angeles Garc\'{\i}a Ferrero  \\
 Instituto de Ciencias Matem\'aticas, C/ Nicol\'as Cabrera, 13-15, 28049
Madrid, Spain
        }

\abstract{
How much information is added to the Review of Particle
Physics when a new decay branching ratio of a hadron is measured and
reported? This is quantifiable by Shannon's information entropy 
$S(i) =-\sum_{f}BR_{(i\longrightarrow f)}\log BR_{(i\longrightarrow f)}$
that takes as input the experimental branching ratios ($BR$) of the decay distribution of particle $i$.  

It may be used at two levels, against the distribution of decay-channel probabilities,
or against the distribution of individual quantum-state probabilities  
(integrating the phase space of those states provides the former). We illustrate the concept
with some examples.
}

\FullConference{EPS-HEP 2017, European Physical Society conference on High Energy Physics\\
		5-12 July 2017\\
		Venice, Italy}

\begin{document}

In this note we deploy Shannon's entropy~\cite{Shannon:1948zz}, a concept belonging to the field of information theory (as a measure of the uncertainty associated with a random variable, or 
on ignoring the value that this takes,  of the average missing information content) to analyze particle decay distributions. This is granted because the sum of branching ratios $BR_i=\Gamma_i/\Gamma$, $i=1\dots N$ is a probability distribution since $\sum_i \Gamma_i=\Gamma$ so that $\sum_i BR_i= 1$. We exemplify with actual data from meson and gauge boson decays, all taken from~\cite{Patrignani:2016xqp}.

Often part of the decaying particle width is unaccounted for the known decay channels. We may then assign that unknown width to one last channel, and Shannon's entropy for the decay reads
\begin{equation}
S(i)=-\sum_{f_{\rm known}}BR_{(i\longrightarrow f)}\log_{k}BR_{(i\longrightarrow f)}
-(1-\sum_{f_{\rm known}}BR_{(i\longrightarrow f)})\log_{k}(1-\sum_{f_{\rm known}}BR_{(i\longrightarrow f)}) \ .
\end{equation}
A more sophisticated treatment could go along the lines of the coarse-graining methods of~\cite{Alonso-Serrano:2017poc}.

\begin{figure}
\includegraphics[width=0.45\textwidth]{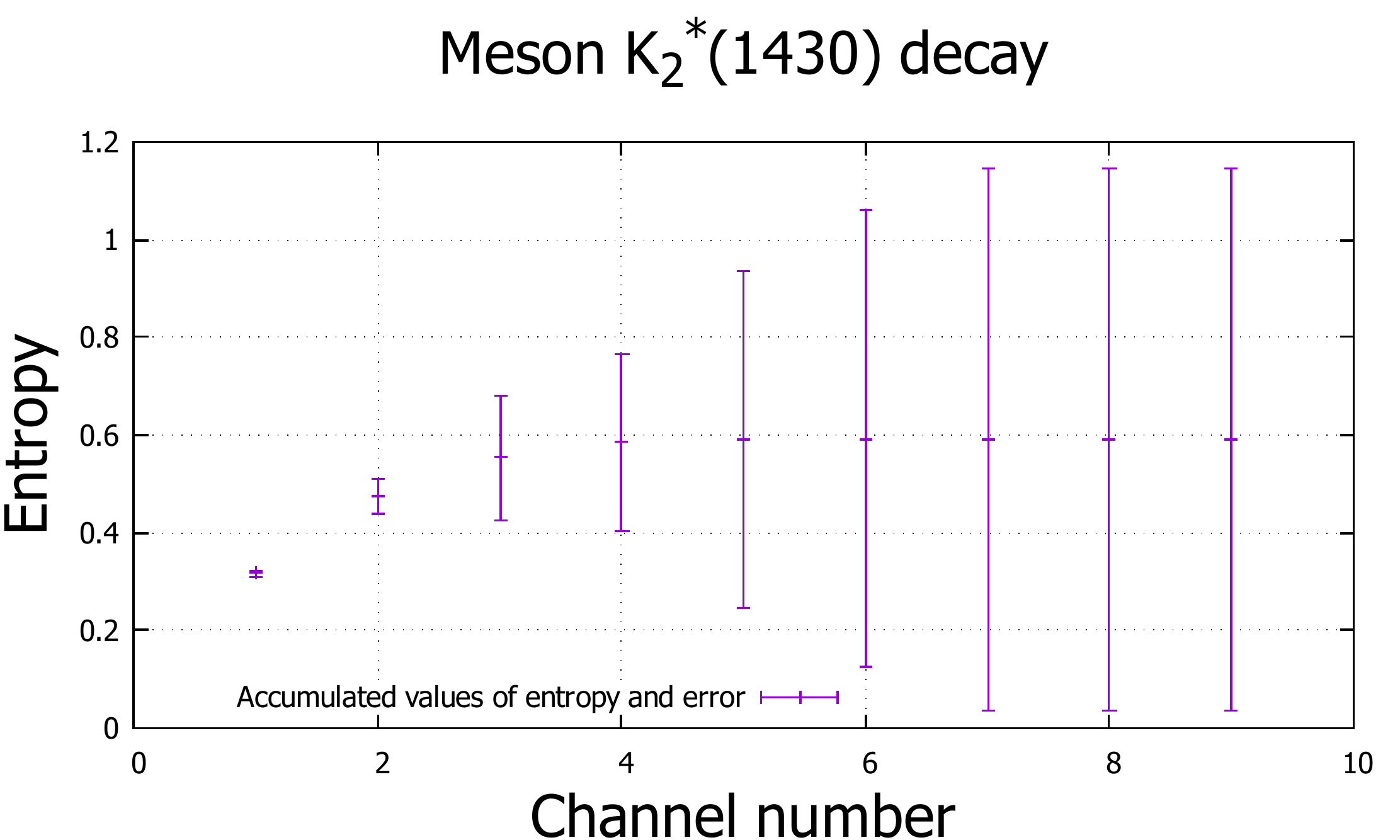} \ 
\includegraphics[width=0.45\textwidth]{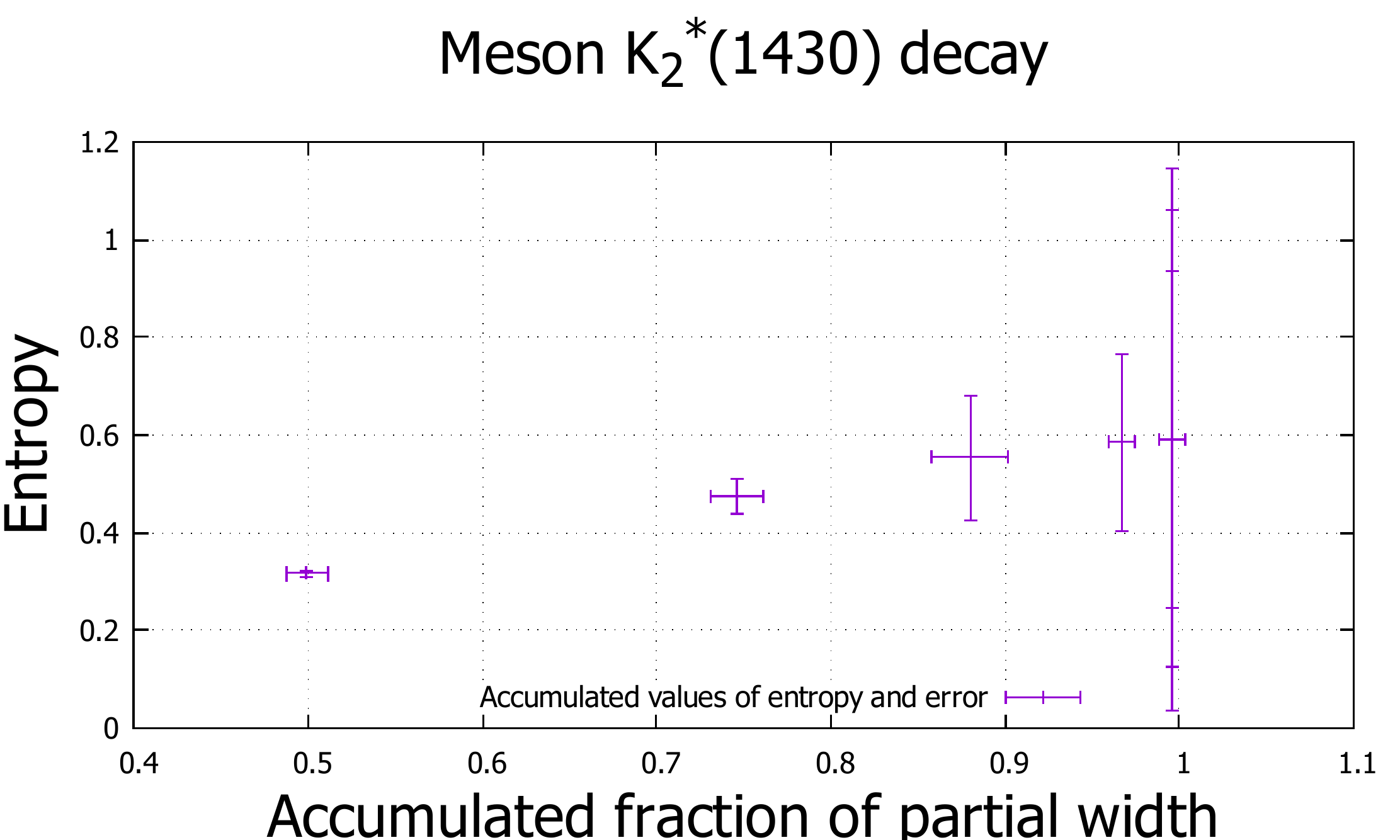} 
\\
\caption{Entropy of the decay distribution of the strange $K_{2}^{*}(1430)$ meson as function of the number of included channels (left) or the accumulated fraction of the partial width after accounting for each channel in order of decreasing branching fraction. The error bars in this and the following figures are propagated from the PDG uncertainties in the measurement of $\Gamma_i/\Gamma$.\label{fig:K2}}
\end{figure}

Figure~\ref{fig:K2} represents the entropy accrued, from left to right, upon adding each decay channel of the $K_2^*(1430)$ meson.
The entropy saturates after the channels with largest branching fraction have been incorporated,
indicating that additional channels, of small branching fraction, barely add any information to the decay distribution.

\begin{figure}[h]
\begin{center}
\includegraphics[width=0.45\textwidth]{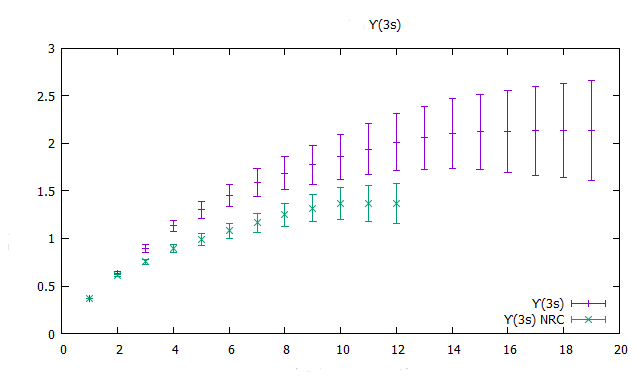}
\caption{Shannon entropy of the decay distribution of  $\Upsilon (3s)$. We compare the result of including all channels with a calculation that ignores radiative ones (denoted NRC, Non-Radiative Channels).\label{fig:upsilon}}
\end{center}
\end{figure}

\begin{figure}[h]
\begin{center}
\includegraphics[width=0.47\textwidth]{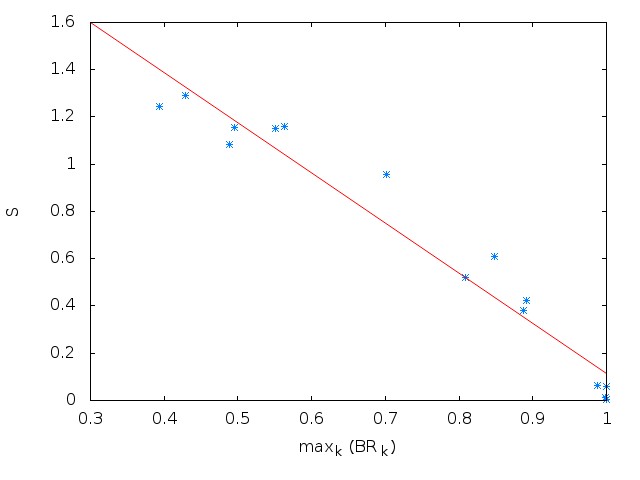}
\caption{This scatter plot, where each point coresponds to a light meson, shows a clear correlation between the entropy $S$ for a decay distribution
and the maximum branching fraction among the channels through which each of the particles can decay.
 \label{fig:correlation}}
\end{center}
\end{figure}

Figure~\ref{fig:upsilon} shows the effect of including or not the $\gamma$-channels of the $\Upsilon(3S)$ decays. It is immediately visible that increasing the number of channels increases the entropy. In fact, the \emph{maximum possible} entropy of a decay distribution with $N$ channels is $\log N$~\cite{Inprep}.
Additionally, we find that a determining parameter that influences the total entropy of a decay distribution is the largest branching fraction of all possible decay channels. This strong correlation is shown in figure~\ref{fig:correlation}.

\begin{figure}
\begin{center}
\includegraphics[width=0.45\textwidth]{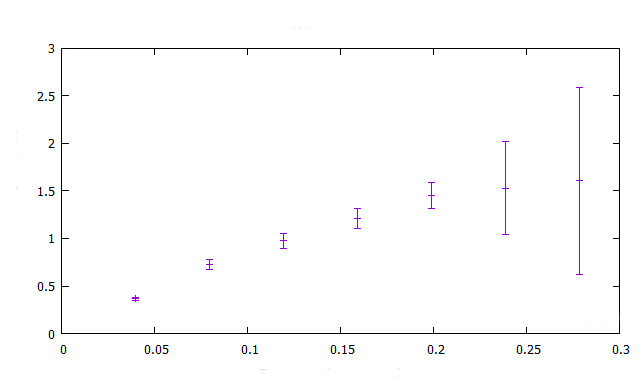}
\caption{Entropy against the phase space $\sum_i \rho_i$ accrued upon including further two--body decay channels of the electroweak $W$ boson, taken from larger to smaller branching fraction.
\label{fig:W}}
\end{center}
\end{figure}

An advantage of employing the entropy with its logarithm of the branching fraction is its \emph{additivity} property upon subdividing a channel into subchannels. For example, one can think of
grouping all channels including a pion into a semiinclusive channel, and later examine them exclusively one by one. Then, $S({\rm both}) = w S(1) + (1-w) S(2)$ where $w$ is the weight of channel 1, that is, $w=BR(1)/(BR(1)+BR(2))$ in terms of the branching ratios. The smallest conceivable subdivision of the decay tree is at the level of individual final quantum states. A practical way of counting these is with the two-body phase space, and figure~\ref{fig:W} shows
the entropy against phase space for all two-body and quasi-two-body decays of the electroweak $W$ boson.

\begin{figure}[h]
\begin{center}
\includegraphics[width=0.45\columnwidth]{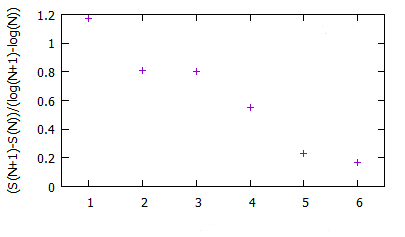} \ 
\includegraphics[width=0.45\columnwidth]{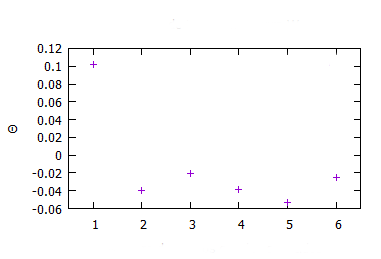}
\caption{ {\bf Left}: Increase of the normalized entropy 
for the decay distribution of the $W$ boson as function of the number of channels. Additional ones can be seen to contribute decreasingly less, so this quantity could be taken as a measure of the amount of information contained in each newly reported channel.
{\bf Right}: Change in the degree of likeness $\Theta$ upon discovering each new channel (so that the last, unknown channel, splits off part of its probability to that new one, and $N$ increases by one unit) for the $W$ boson.
 }\label{fig:normincrement}
\end{center}
\end{figure}

We wish to propose  a simple criterion to quantify what information the discovery of a new branching fraction provides. The first obvious effect is that, since the maximum entropy grows as the logarithm of the number of channels, if all were equally weighted, the actual importance of a new channel can be obtained by studying the separation of the entropy from this maximum value. 
Therefore, we propose two measures of this added information. One is the normalized entropy increment, that is plotted in figure~\ref{fig:normincrement}, as defined by
\begin{equation} \label{normincrement}
\frac{\Delta S(N)}{\Delta log(N)}=\frac{S(N+1)-S(N)}{log(N+1)-log(N)}\ .
\end{equation}

Another possibility is to employ a certain ``degree of likeness'' which can be simply given by
$\frac{S(N)}{log(N)}\in (0,1)$. Its increment upon adding one new channel would then be
\begin{equation}
\Theta = \frac{S(N+1)}{log(N+1)}-\frac{S(N)}{log(N)}\ .
\end{equation}
A positive $\Theta$ says that the entropy of distribution gets closer to the maximum possible upon introducing the new channel; therefore, this new channel has a branching fraction similar to the others. If $\Theta$ is negative, the entropy decreases relative to its maximum possible value, and
the new channel is very dissimilar from the others. This function is also plotted in figure~\ref{fig:normincrement}.

In conclusion, Shannon's entropy is a promising way of assessing and quantifying  the information gained upon discovering a new decay channel of a particle.


\end{document}